\documentclass{elsart}

\newcommand{\lag}{{\mathcal{L}}}
\newcommand{\ut}{\tilde{u}}
\newcommand{\dt}{\tilde{d}}
\newcommand{\ltsim}{\lower3pt\hbox{$\, \buildrel < \over \sim \, $}}
\newcommand{\gtsim}{\lower3pt\hbox{$\, \buildrel > \over \sim \, $}}

\usepackage{amssymb,graphics}

\begin{document}

\begin{frontmatter}

\rightline{UG--FT--119/00}
\rightline{hep-ph/0008143}
\rightline{August 2000}



\title{Universality limits on bulk fermions}


\author{F. del  Aguila\thanksref{faguila}} and
\author{J. Santiago\thanksref{jsantiag}}

\address{Departamento de F\'{\i}sica Te\'{o}rica y del Cosmos \\
Universidad de Granada \\
E-18071 Granada, Spain}
\thanks[faguila]{faguila@ugr.es}
\thanks[jsantiag]{jsantiag@ugr.es}

\date{\today}

\begin{abstract}
The Kaluza-Klein fermion excitations induce mixing between the
Standard Model fermions and loss of
universality. The flavour mixing not present in the Standard Model 
 can be made to vanish aligning the
Yukawa couplings and the Dirac masses of the heavy modes, 
but universality is only recovered
when these masses go to infinity. This implies a bound
 on the lightest new heavy quark, $M_1
\gtsim 3-5$ TeV, which together with the electroweak precision data
limits will allow the Large Hadron Collider to provide a crucial test
of the Randall-Sundrum ansatz for solving the
gauge hierarchy.
\end{abstract}

\end{frontmatter}

Attempts to solve the gauge hierarchy problem using models with extra
dimensions have received a great deal of attention during the past few
years~\cite{antoniadis}. In a proposal by Arkani-Hamed, Dimopoulos and
Dvali  the  hierarchy between the Planck
and electroweak scales is related to the large volume of the extra
dimensions where only gravity propagates~\cite{aad}. 
Randall and Sundrum (RS) proposed an alternative solution based on a
non-factorizable geometry with a warped background metric in
a slice of ${\rm AdS}_5$~\cite{rs}. 
The exponential warp factor, obtained
imposing four-dimensional Poincar\'{e} invariance, accounts in this
case  for the hierarchy between the
Planck and the electroweak  scales (see however Ref.~\cite{strings}).
These models with extra dimensions also give
 a rich TeV phenomenology. In the RS proposal,
only gravity is assumed to propagate in the extra dimension. 
However the possibility of placing the Standard Model (SM) fields  in the
five-dimensional bulk has been also considered in the literature. 
Bulk scalars were analysed in Ref.~\cite{goldberger}, gauge
bosons propagating in the bulk were studied in
Ref.~\cite{gb}, and fermions were included for the
first time in Ref.~\cite{grossman}. The complete SM living in the
bulk has been treated in Ref.~\cite{chang:SM} and a 
complete parametrization of bulk field masses and their
phenomenology can be found in Ref.~\cite{gherghetta}, where supersymmetry
is also discussed. Finally, the  phenomenology  
of the RS model with the fields living on and off the wall, 
is reviewed in Ref.~\cite{davoudiasl:all},
where experimental bounds and the reach
 of Tevatron and the Large Hadron Collider (LHC) are studied in
detail. 

Fields living in the five-dimensional bulk can be expanded as a tower of
Kaluza-Klein (KK) four-dimensional states, with the mass of each level
being the (quantized) momentum in the transverse dimension. In the
case of SM fermions, chiral zero modes can be obtained using an
orbifold projection in the fifth dimension.
The rest of the KK tower, however, is
 necessarily vector-like: the massive modes are Dirac particles whose
Left-Handed (LH) and Right-Handed (RH) parts transform in the same way
under the SM gauge group. The presence of these extra
fermions in the spectrum
can induce large mixing between the SM zero modes. 
This possibility has
been usually neglected arguing the mixing suppression 
due to the large KK fermion masses $M$.
Indeed experiment tends to banish these KK excitations above  
$\sim \mathrm{TeV}$, implying
at least a suppression of  
$\frac{v^2}{M^2}\sim\left(\frac{0.25
\mathrm{TeV}}{1 \mathrm{TeV}} \right)^2\sim 0.06$, 
where $v$ is the electroweak vacuum espectation value. However, 
smaller $M$ masses are also possible. For instance, 
it can be shown in the RS model with fermions in the bulk 
that there is a point in
parameter space where the conformal limit is
recovered~\cite{gherghetta}. The
five-dimensional momentum is then 
conserved and couplings between zero
mode fermions and a non-zero mode gauge boson are forbidden. As a
result,
 bounds on the new gauge boson masses coming from electroweak precision
data or direct production of KK gauge boson excitations  do not apply.
 Moreover, around this point 
fermion couplings  to the graviton
tower remain small.
 In summary, the KK excitations could have masses 
 $\sim 0.5$ TeV or even smaller in this region of parameter space,  
depending on the ratio
between the bulk curvature and the Planck mass~\cite{davoudiasl:all}. 
Thus, ignoring  
quark mixing and the corresponding universality constraints, the
experimental information available at present and even after LHC    
 leaves an open window to small KK masses.
In this paper we discuss the constraints on the mixing induced by
 bulk fermion excitations. 
Their contributions to fermion couplings can be readily 
read from Ref.~\cite{apvs}. 
There the SM extension with an arbitrary number of 
vector-like fermions is considered and the effective 
Lagrangian resulting from integrating them out obtained. 
Using this Lagrangian we show in the following 
that to keep the quark mixing in the RS model  
small enough to fulfil universality to few 
per cent, the lightest KK quark excitation 
must be heavier than $3-5$ TeV in the window around the conformal point. 
Vector-like quark masses will be also constrained by direct
production at LHC.
As a matter of fact, a lower bound of $\sim 1.5$ TeV will be 
placed on them if none of such quarks is observed~\cite{delAguila}.  
However, the universality limit derived from the 
expected precision in the determination of the top 
couplings at LHC~\cite{Beneke:2000hk} is  
larger by more than a factor of 2. 
This will close the window of small masses, 
allowing LHC to test crucially the RS
ansazt for solving the gauge hierarchy problem. We assume
the consistency of this model with SM fields off the
boundary. It will be shown that all the mixing contributions 
of the KK fermions have the same sign, 
with the total sum being dominated by the first excited states. 
It seems improbable that other contributions to
two-fermion gauge couplings cancel those from KK fermion excitations,
which is the largest tree level source for mixing beyond the
SM~\cite{delAguila:2000aa}. 

Let us first review the RS model to fix our notation~\cite{rs}.
The topology of the fifth dimension is an orbifold
 $\mathrm{S}_1/\mathrm{Z}_2$ of
radius R, with two 3-branes sitting on the orbifold fixed points
($y=0$ for the Planck boundary and $y=\pi R$ for the TeV boundary). 
The background metric which satisfies five-dimensional 
Einstein's equations and four-dimensional Poincar\'{e} invariance
reads  $\d s^2=\e^{-2\sigma} \eta_{\mu\nu} \d x^\mu \d
x^\nu+\d y^2$, with $\sigma=k|y|$ and $1/k$ the ${\rm AdS}_5$ curvature
radius. The exponential warp factor in the metric reduces
the only fundamental scale, the Planck mass 
$M_{Pl}\sim \mathcal{O}(k)$, to TeV
scales on the  $y=\pi R$ boundary,
$M_{Pl}\e^{-\pi k R}\sim \mathcal{O}$(TeV) provided 
$k R\sim \mathcal{O}(10)$. 
For simplicity we assume that the SM Higgs lives on the TeV boundary, as
it is phenomenologically 
preferred~\cite{chang:SM,gherghetta,davoudiasl:all,Kitano,huber}. 
We consider that the SM fermions live in the bulk since we want to
study the effects of their KK excitations. The SM gauge
bosons are also allowed 
to propagate in the fifth dimension to maintain in general gauge
invariance.  The KK excitations of the gauge
bosons can also introduce fermion mixing but this has been discussed
elsewhere~\cite{Delgado:2000sv}, and we will not study this
possibility  in detail here.

In five dimensions there are no chiral fermions. Thus five-dimensional 
fermions $\Psi$ are vector-like and can have a Dirac mass of the form 
\begin{equation}
\lag_\mathrm{D}=- i m_\Psi (\bar{\Psi}_\mathrm{L} \Psi_\mathrm{R} +
\bar{\Psi}_\mathrm{R} \Psi_\mathrm{L}),
\label{lag:dirac}\end{equation}
where $\Psi$ is the sum of the two four-dimensional ``chiralities'' 
$\Psi_{\mathrm{L,R}}=\pm \gamma_5 \Psi_{\mathrm{L,R}}$, 
transforming in the same way under the gauge group.
However, in the RS background $\Psi_\mathrm{L}$ and $\Psi_\mathrm{R}$
must have opposite parities under the $\mathrm{Z}_2$ symmetry $y
\to -y$. 
This implies that the Dirac mass must present a kink profile
and can be parametrized as $m_\Psi=c_\Psi \sigma^\prime$, where
$c_\Psi$ 
is a free parameter determining the location of the
zero mode~\cite{gherghetta}. 
This symmetry can be used to classify the chirality of the four
dimensional states, surviving as zero modes only those with even
chirality.
 In this way a massless chiral spectrum can be generated.
The KK expansion of the fermion fields can be written
~\cite{grossman,gherghetta}
\begin{equation}
\Psi_{L,R}(x^\mu,y)=\frac{\e^{2\sigma}}{\sqrt{2\pi R}}\sum_{n=0}^\infty
\Psi_{L,R}^{(n)}(x^\mu)f^{L,R}_n(y),
\end{equation}
where the expansion coefficients depending on
the coordinate transverse to the brane read for the L and R projections
 ($n\neq 0$)
\begin{equation}
f_n(y)=\frac{\e^{\sigma/2}}{N_n}
\left[
J_\alpha(\frac{M_n}{k}\e^\sigma) +b_\alpha(M_n)
Y_\alpha(\frac{M_n}{k}\e^\sigma) \right].
\end{equation}
 $M_n>0$ is the mass of the KK excitation $\Psi^{(n)}$; 
$J_\alpha$ and
$Y_\alpha$ are Bessel functions of order
$\alpha=|c\pm\frac{1}{2}|$, where the signs $\pm$ correspond 
to $f^{L,R}$, respectively; and 
the normalization constants 
\begin{equation}
N^2_n=\frac{1}{\pi R}\int^{\pi R}_0 \d y\; \e^{2\sigma}  
\left[
J_\alpha(\frac{M_n}{k}\e^\sigma) +b_\alpha(M_n)
Y_\alpha(\frac{M_n}{k}\e^\sigma) \right]^2 ,
\end{equation}
take in the limit $M_n\ll k$ and $kR\gg 1$ the approximate form
\begin{equation}
N_n\simeq \frac{\e^{\pi k R/2}}{\sqrt{\pi^2 R\, M_n}}.
\end{equation}
The constant coefficients $b_\alpha(M_n)$ together with the KK
masses are calculated using boundary conditions.
For even fields
\begin{equation}
b_\alpha(M_n)=-\frac{(\pm c+1/2)J_\alpha(M_n/k)+\frac{M_n}{k}
J^\prime_\alpha(M_n/k)}  {(\pm c+1/2)Y_\alpha(M_n/k)+\frac{M_n}{k}
Y^\prime_\alpha(M_n/k)},
\end{equation}
and
\begin{equation}
b_\alpha(M_n)=b_\alpha(M_n \e^{\pi k R}),
\end{equation}
with $\pm$ for $f^{L,R}$, respectively. In the limit 
$M_n\ll k$ and $kR\gg 1$,
\begin{equation}
M_n\simeq (n+\frac{\alpha}{2}-\frac{3}{4}) \pi k \e^{-\pi k R}.
\end{equation}

We omit the corresponding equations for odd fields because we will not
use them explicitly here, for odd boundary
conditions do not allow  
massless zero-mode solutions. Besides, their massive
modes do not couple to the boundary and they cannot acquire masses
through Yukawa couplings with a boundary Higgs.
The coefficients of the zero modes for
the even chiralities are
\begin{equation}
f_0^{\mathrm{L,R}}(y)=\frac{\e^{\mp c\sigma(y)}}{\sqrt{\frac{\e^{(1\mp
2c)\pi k R}-1}{(1\mp 2c) \pi k R}}}.
\end{equation}
Although we are considering fields living in the five-dimensional
bulk, the ${\rm AdS}_5$ space can localize the different KK states.
The value of the mass parameter $c$ determines the location of the
zero mode. When $c_{L(R)}=\frac{1}{2} \; (-\frac{1}{2})$ the conformal
limit is recovered, 
the kinetic terms are independent of $y$ 
and the five-dimensional momentum is conserved.
 In this case the zero mode is flat. Values of
$c_{L(R)}$  greater (smaller) 
than $\frac{1}{2}$ ($-\frac{1}{2}$) localize the zero mode near
the Planck boundary, while values $c_{L(R)}<\frac{1}{2}$
($>-\frac{1}{2}$) imply that the zero mode is localized near the
TeV boundary. The value of the trilinear couplings
between the fermion zero modes and the tower of KK gauge bosons also
depends on the value of $c$, being zero for $c_{L(R)}=\frac{1}{2}$
($-\frac{1}{2}$) and
essentially constant for $c_{L(R)}>\frac{1}{2}$ 
($<-\frac{1}{2}$)~\cite{gherghetta}. The
non-zero modes are always localized near the TeV brane.

In order to reproduce the SM we consider 
three quark doublets $q_i$ with even LH 
parts, three up-type singlets $\tilde{u}_i$ 
with even RH parts and
 three down-type quark singlets $\tilde{d}_i$ also with even
RH parts. 
(We use a tilde for the singlets to better distinguish in the following
the tower of KK states.)
Their opposite chiralities are odd. All quarks
live in the bulk and have mass parameters $c^q_i,c^u_i$, and
$c^d_i$,  respectively.  
The five-dimensional action containing the 
Yukawa interactions can  be written in general 
\begin{eqnarray}
S_\mathrm{Yuk}&=&-i
\int \d^4x\;\int \d y\;\sqrt{-g}\left[ \lambda^{u(5)}_{ij} \bar{q}_i(x,y)
\tilde{u}_j(x,y) \tilde{\phi}(x)   \right. \nonumber \\
&&+ \left.  \lambda^{d(5)}_{ij} \bar{q}_i(x,y)
\tilde{d}_j(x,y) \phi(x) + \mathrm{h.c.}
\right] \delta(y-\pi R).\label{S:yuk}
\end{eqnarray} 
Expanding the five-dimensional fields in KK towers and 
integrating over the fifth dimension, we find 
after spontaneous symmetry breaking a
four-dimensional mass Lagrangian of the form
\begin{eqnarray}
i\lag_\mathrm{mass}&=&\sum_{n,m=0}^\infty\left[ 
\lambda^{u(nm)}_{ij}
\bar{u}^{(n)i}_L\ut^{(m)j}_R
+\lambda^{d(nm)}_{ij}
\bar{d}^{(n)i}_L\dt^{(m)j}_R
\right] +\mathrm{h.c.} \nonumber \\
&& +\sum_{n=0}^\infty \left[
M^{q(n)}_i
( \bar{u}^{(n)i}_L u^{(n)i}_R+\bar{u}^{(n)i}_R u^{(n)i}_L+
\bar{d}^{(n)i}_L d^{(n)i}_R+\bar{d}^{(n)i}_R d^{(n)i}_L) \right.
 \nonumber \\
&&\quad \quad
+M^{u(n)}_i (\bar{\ut}^{(n)i}_L \ut^{(n)i}_R +\bar{\ut}^{(n)i}_R
\ut^{(n)i}_L) \nonumber \\
&& \quad \quad
\left.+M^{d(n)}_i (\bar{\tilde{d}}^{(n)i}_L \dt^{(n)i}_R +\bar{\dt}^{(n)i}_R
\dt^{(n)i}_L)\right],\label{lag:mass}
\end{eqnarray}
where we have added to Eq. (\ref{S:yuk}) the Dirac masses in Eq.
(\ref{lag:dirac}). These can always be taken diagonal.
 The four-dimensional Yukawa couplings are
\begin{eqnarray}
\lambda^{u(nm)}_{ij}&=&\lambda^{u(5)}_{ij} \frac{v}{\sqrt{2}}\e^{\pi k R}
\frac{f_{qL}^{(n)i}(\pi R) f_{uR}^{(m)j}(\pi R)}{2\pi
R}\equiv\lambda^u_{ij} a_q^{(n)i} a_u^{(m)j}, \\
\lambda^{d(nm)}_{ij}&=&\lambda^{d(5)}_{ij} \frac{v}{\sqrt{2}}\e^{\pi k R}
\frac{f_{qL}^{(n)i}(\pi R) f_{dR}^{(m)j}(\pi R)}{2\pi
R}\equiv\lambda^d_{ij} a_q^{(n)i} a_d^{(m)j}, 
\end{eqnarray}
with
\begin{eqnarray}
v&=&\e^{-\pi k R} v^{(5)}\sim 250 \; \mathrm{GeV}, \\
\lambda^{u,d}_{ij}&=&\lambda^{u,d(5)}_{ij} k \frac{v}{\sqrt{2}}
\sim \mathrm{\;SM\; masses}, \\
a_q^{(n)i} &=& \e^{\pi k R/2} \frac{f_{qL}^{(n)i}(\pi R)}{\sqrt{2\pi k
R}}, \\
a_{u,d}^{(m)j}&=& \e^{\pi k R/2} \frac{f_{u,dL}^{(m)j}(\pi
R)}{\sqrt{2\pi k R}}. 
\end{eqnarray}
Notice that the factor
 $\e^{\pi k R}$ in the definition of $\lambda^{u,d(nm)}_{ij}$
is due to the rescaling of the boundary Higgs 
canonically normalized. Note also that 
 odd fields are zero at the TeV boundary and then the odd
chiralities ($q_\mathrm{R},\tilde{u}_\mathrm{L},\tilde{d}_\mathrm{L}
$) have zero Yukawa couplings for a boundary Higgs.
In matrix notation Eq. (\ref{lag:mass}) reads
\begin{equation}
{\mathcal{M}}^u=
\begin{array}{l}
 \\ \bar{u}^{(0)}_L \\\bar{\ut}^{(1)}_L   \\ \vdots \\
\bar{u}^{(1)}_L \\ \vdots
\end{array}
\begin{array}{c} \quad \quad 
  \ut^{(0)}_R  \quad \quad
\ut^{(1)}_R  \quad \quad \quad \ldots  \quad
u^{(1)}_R  \quad  \ldots \\
\left(
\begin{array}{ccccc}
 \lambda^u_{ij} a^{(0)i}_q a^{(0)j}_u 
 & \lambda^u_{ij} a^{(0)i}_q a^{(1)j}_u 
&  \ldots & 0 & \dots \\
 0 & M^{u(1)}_i \delta_{ij}
 & \ldots & 0  & \dots \\
 \vdots & \vdots
 & \ddots & \vdots &  \\
 \lambda^u_{ij} a^{(1)i}_q a^{(0)j}_u  & \lambda^u_{ij} a^{(1)i}_q
 a^{(1)j}_u   
 & \ldots & M^{q(1)}_i \delta_{ij}  & \dots \\
 \vdots & \vdots &  & \vdots  & \ddots
\end{array}\right) , \end{array}
\end{equation}
and similarly for ${\mathcal{M}}^d$. 

In order to obtain the effective Lagrangian describing the 
interactions between the SM quarks, we integrate out 
the heavy quark excitations. This has been done for generic 
vector-like quark additions in Ref.~\cite{apvs}. 
To use the results there, we must first rotate the zero modes,
\begin{equation}
\tilde{u}^{(0)i}_R=(U^u_R)_{ij} \tilde{u}^{\prime(0)j}_R, \quad 
\tilde{d}^{(0)i}_R=(U^d_R)_{ij} \tilde{d}^{\prime(0)j}_R,\quad 
q^{(0)i}_L=(U^q_L)_{ij} q^{\prime(0)j}_L,
\end{equation}
such that 
\begin{equation}
(U^{q\dagger}_{L})_{ik} \lambda^u_{kl} a^{(0)k}_q a^{(0)l}_u
(U^u_R)_{lj} 
=
V^\dagger_{ij} m^u_j,\quad
(U^{q\dagger}_{L})_{ik} \lambda^d_{kl} a^{(0)k}_q a^{(0)l}_d 
(U^d_R)_{lj}  =
m^d_i \delta_{ij}. \label{rotations}
\end{equation}
In the SM $m^{u,d}_i$ are the quark masses and $V$ 
the Cabibbo-Kobayashi-Maskawa  (CKM) matrix. To order $M^{-2}$  
the quark couplings to $Z$ and $W^{\pm}$, $X^{u,dL,R}_{ij}$ and 
$W^{L,R}_{ij}$, respectively, are in the mass eigenstate basis
~\cite{apvs} 
(sums on family indices are understood throughout
the paper)
\begin{eqnarray}
X^{uL}_{ij}&=&\delta_{ij}-
\sum_{n=1}^\infty 
\frac{m^{\prime(1)\dagger}_{i,nk}
m^{\prime(1)}_{nk,j}}{M^{u(n)\;2}_k},\label{xul} \\
X^{uR}_{ij}&=&
\sum_{n=1}^\infty 
\frac{m^{\prime(3u)\dagger}_{i,nk}
m^{\prime(3u)}_{nk,j}}{M^{q(n)\;2}_k}, \label{xur} \\
X^{dL}_{ij}&=&\delta_{ij}-
\sum_{n=1}^\infty 
\frac{m^{\prime(2)\dagger}_{i,nk}
m^{\prime(2)}_{nk,j}}{M^{d(n)\;2}_k}, \label{xdl} \\
X^{dR}_{ij}&=&
\sum_{n=1}^\infty 
\frac{m^{\prime(3d)\dagger}_{i,nk}
m^{\prime(3d)}_{nk,j}}{M^{q(n)\;2}_k}, \label{xdr} \\
W^{L}_{ij}&=&\tilde{V}_{ij}
-\frac{1}{2}
\sum_{n=1}^\infty 
\frac{m^{\prime(1)\dagger}_{i,nk}
m^{\prime(1)}_{nk,l}}{M^{u(n)\;2}_k}
\tilde{V}_{lj}
-\frac{1}{2}
\tilde{V}_{il}
\sum_{n=1}^\infty 
\frac{m^{\prime(2)\dagger}_{l,nk}
m^{\prime(2)}_{nk,j}}{M^{d(n)\;2}_k},\label{wl} 
\\
W^{R}_{ij}&=&
\sum_{n=1}^\infty 
\frac{m^{\prime(3u)\dagger}_{i,nk}
m^{\prime(3d)}_{nk,j}}{M^{q(n)\;2}_k},\label{wr} 
\end{eqnarray}
where $\tilde{V}$ is the corrected unitary CKM matrix~\cite{apvs}. 
 The superscripts $(1)$,$(2)$, and
$(3u)$,$(3d)$ stand for mixing with heavy
vector-like up and down singlets, and up and down quarks within doublets,
respectively. We have introduced the definitions
\begin{eqnarray}
m^{\prime(1)}_{nk,j}&\equiv&\lambda^{u\dagger}_{k l}
a^{(0)l}_q a^{(n)k}_u
 (U^q_L)_{lr}V^\dagger_{rj}=
\frac{a^{(n)k}_u}{a^{(0)k}_u} (U^u_R)_{kj} m^u_j,\label{mp1} 
\\
m^{\prime(2)}_{nk,j}&\equiv&\lambda^{d\dagger}_{k l}
a^{(0)l}_q a^{(n)k}_d
 (U^q_L)_{lj}=\frac{a^{(n)k}_d}{a^{(0)k}_d} (U^d_R)_{kj}
m^d_j,\label{mp2}
  \\
m^{\prime(3u)}_{nk,j}&\equiv&\lambda^{u}_{k l} a^{(n)k}_q a^{(0)l}_u
(U^u_R)_{lj}=\frac{a^{(n)k}_q}{a^{(0)k}_q} (U^q_L)_{kl} V^\dagger_{lj} 
m^u_j,\label{mp3u}  \\
m^{\prime(3d)}_{nk,j}&\equiv&\lambda^{d}_{k l}a^{(n)k}_q a^{(0)l}_d
 (U^d_R)_{lj}=\frac{a^{(n)k}_q}{a^{(0)k}_q} (U^q_L)_{kj}
m^d_j,\label{mp3d}   
\end{eqnarray}
where the second equalities follow from Eq. (\ref{rotations}).
These allow to rewrite the
$X_{ij}$ and $W_{ij}$ corrections 
in a more transparent way to analyse flavour
mixing
\begin{eqnarray}
X^{uL}_{ij}&=&\delta_{ij}-
m^u_i (U^{u\dagger}_R)_{ik}
\left[\sum_{n=1}^\infty 
\left(\frac{a^{(n)k}_u}{a^{(0)k}_u}\right)^2 
\frac{1}{M^{u(n)\;2}_k}
\right]
 (U^{u}_R)_{kj} m^u_j , 
\label{xul1} \\
X^{uR}_{ij}&=&
m^u_i V_{il} 
(U^{q\dagger}_L)_{lk}
\left[\sum_{n=1}^\infty 
\left(\frac{a^{(n)k}_q}{a^{(0)k}_q}\right)^2 
\frac{1}{M^{q(n)\;2}_k}
\right]
 (U^{q}_L)_{kr} V^\dagger_{rj} m^u_j , 
\label{xur1} \\
X^{dL}_{ij}&=&\delta_{ij}-
m^d_i (U^{d\dagger}_R)_{ik}
\left[\sum_{n=1}^\infty 
\left(\frac{a^{(n)k}_d}{a^{(0)k}_d}\right)^2 
\frac{1}{M^{d(n)\;2}_k}
\right]
 (U^{d}_R)_{kj} m^d_j ,
\label{xdl1}  \\
X^{dR}_{ij}&=&
m^d_i  (U^{q\dagger}_L)_{ik}
\left[\sum_{n=1}^\infty 
\left(\frac{a^{(n)k}_q}{a^{(0)k}_q}\right)^2 
\frac{1}{M^{q(n)\;2}_k}
\right]
 (U^{q}_L)_{kj}  m^d_j , 
\label{xdr1} \\
W^{L}_{ij}&=&\tilde{V}_{ij}
-\frac{1}{2}
m^u_i (U^{u\dagger}_R)_{ik}
\left[\sum_{n=1}^\infty 
\left(\frac{a^{(n)k}_u}{a^{(0)k}_u}\right)^2 
\frac{1}{M^{u(n)\;2}_k}
\right]
 (U^{u}_R)_{kl} m^u_l
\tilde{V}_{lj} \nonumber \\
&&-\frac{1}{2}
\tilde{V}_{il}
m^d_l (U^{d\dagger}_R)_{lk}
\left[\sum_{n=1}^\infty 
\left(\frac{a^{(n)k}_d}{a^{(0)k}_d}\right)^2 
\frac{1}{M^{d(n)\;2}_k}
\right]
 (U^{d}_R)_{kj} m^d_j , 
\label{wl1} \\
W^{R}_{ij}&=&
m^u_i V_{il} (U^{q\dagger}_L)_{lk}
\left[\sum_{n=1}^\infty 
\left(\frac{a^{(n)k}_q}{a^{(0)k}_q}\right)^2 
\frac{1}{M^{q(n)\;2}_k}
\right]
 (U^{q}_L)_{kj}  m^d_j , 
\label{wr1}
\end{eqnarray}
where at this order $V$ can be replaced by $\tilde V$ in 
$X^{uR}$ and $W^R$.
 The new contributions correcting the SM values,  
$X^{u,dL}_{ij}=\delta_{ij}$, $X^{u,dR}_{ij}=0$, 
$W^{L}_{ij}=V_{ij}$, $W^{R}_{ij}=0$, are 
products of three $3\times 3$ matrices, where the second one in 
square brackets is diagonal and the entries of the other two are 
combinations of SM masses.  
The matrix in the middle can be further simplified 
noting that
\begin{equation}
a^{(n)}=(-1)^{n-1}a^{(1)}, 
\end{equation}
which, up to a constant, leaves the diagonal elements 
as an infinite sum of the inverse of the KK heavy masses squared.  
As can be observed from 
Eqs.~(\ref{xul1}-\ref{xdr1}), there are only two ways to 
ensure the absence of Flavour Changing Neutral Currents (FCNC),
that is to have only diagonal $X$ corrections. One is 
 that the Yukawa couplings are aligned with the
Dirac masses, in which case the
rotation matrices $U$ are equal to the identity. The other, that the terms 
in square brackets are proportional to the identity, which can be only 
accomplished if all the fields of the same type 
are located at the same point. This means that 
each type of quark has a common, flavour independent 
mass parameter, $c^{q,u,d}_i=c^{q,u,d}$. 

\begin{center}
\begin{figure}[!h]
\caption{$90\%$ C.L. limit on the mass $M_1$ of the first KK quark 
excitation assuming a deviation of $X^{uL}_{tt}$ from its SM value, 1, 
smaller than $5\%$ for $c^q_i=-c^u_i=-c^d_i=c$. The shadowed band is the 
open window when quark mixing is not taken into account.}\label{cotas}
\includegraphics{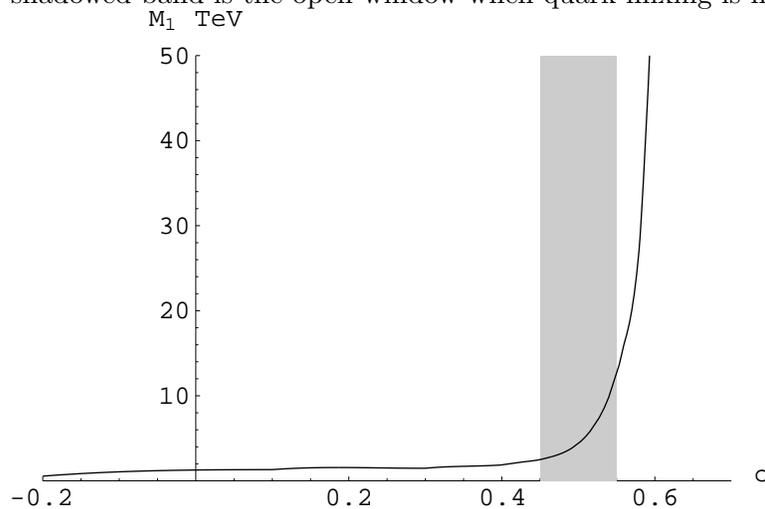}
\end{figure}
\end{center}

Let us discuss first the simplest case with all mass parameters equal, 
$c^q=-c^u=-c^d=c$.
Then the square brackets in Eqs.~(\ref{xul1}-\ref{wr1}) only depend 
on the parameter $c$ for given curvature and warp factor. 
For $c < \frac{1}{2}$ the ratio of constants 
$\frac{a^{(1)}}{a^{(0)}}$ is order 1 and the sum 
is order $\frac{1}{M_1^2}$, the inverse of the squared mass of the 
lightest KK heavy mode. 
 For $c \gtsim \frac{1}{2}$ the ratio grows exponentially (and so does 
$M_1 \propto ke^{-\pi kR}$ if the value of the square bracket 
is to remain constant). 
 This is due to the very small zero mode wave function on the 
TeV boundary~\cite{gherghetta}. 
(The same property allows to obtain small neutrino 
masses in this context \cite{grossman}.) 
Asking for deviations from the SM value 
of the diagonal
top coupling $X^{uL}_{tt}=1$ (which is the quantity
receiving the largest correction for it is proportional to 
$m^2_t$)
smaller than few percent, which is the precision to be reached at 
LHC~\cite{Beneke:2000hk}, 
a limit on the square bracket value 
in Eq. (\ref{xul1}) can be derived. 
This translates into a limit on 
$M_1=M_1^u\simeq 2.45ke^{-\pi kR} $ (where the numerical factor
corresponds to $c=\frac{1}{2}$) as a function of $c$. 
 In Fig.~\ref{cotas} we plot the $90\%$ C.L. $M_1$ bound  
assuming that $X^{uL}_{tt}$ is 
measured at LHC within $5\%$ of its SM value. 
(Notice the change of notation with respect to
Ref.~\cite{davoudiasl:all}. Our results can be compared replacing
$c$ by $-\nu$.)
This shows that the narrow window left open after LHC 
if quark mixing is neglected, $0.45 \ltsim c \ltsim 0.55$
~\cite{davoudiasl:all}, is
completely closed with the precise measurement of the diagonal top 
coupling.
In this region the $M_1$ limit ranges from 2.5 to 12.7
TeV. 
If the  experimental
result coincides within $1\%$ with the SM prediction, the
$M_1$ lower bound will vary from 5.5 to 28.4 TeV in the same region.
Notice that this bound is independent of the ratio $k/M_{Pl}$.
This is not the case for the bounds coming from direct
production of KK graviton states at large colliders. 
In particular, for $k=0.01 M_{Pl}$
the $M_1$ lower limit will be $\gtsim 0.5$ TeV in the region 
$c\sim 0.5$ if no graviton signal is seen~\cite{davoudiasl:all}. 
Direct production of vector-like (KK) quarks will imply 
$M_1 \gtsim 1.5$ TeV if such heavy quarks are not observed 
either~\cite{delAguila}.  

Other indirect constraints not involving the top quark are 
less restrictive. Thus, although  $X^{dL}_{bb}$ has 
been measured with a precision of $0.5\%$
~\cite{delAguila:1999tp,Groom:2000in}, 
the corresponding 
correction in Eq. (\ref{xdl1}) is proportional to $m_b^2$ and 
the bound on $M_1=M_1^d$ is reduced by a factor  
$\sqrt {\frac{0.05}{0.005}}\frac{m_b}{m_t} \sim 0.09 $, 
varying from 0.2 to 1.1 TeV for 0.45 $<c=-c^d<$ 0.55. 
Similarly,  the unitarity 
condition $\sum_{j=1}^3 |W^L_{uj}|^2 = 1$ 
is satisfied to few per mille ~\cite{Groom:2000in}, 
but the new contributions in Eq. (\ref{wl1}) are 
proportional to  
$m_u^2$ and to $\sum_{j=1}^3 |\tilde V _{uj}|^2 m_j^{d\ 2}$, respectively. 
Hence,  
the corresponding limits on 
$M_1^u$ and $M_1^d$ (equal to $M_1$ 
for a unique $c$) are further suppressed by a 
small mass and by small mixing angles times small masses, 
respectively.

If each type of quark has a different location, the same limits 
on $M_1^{u,d}$ above apply. 
(The bound on  $M_1^q$ is equal to the  $M_1^u$ limit if 
$X^{uR}_{tt}$ is measured with the same precision as $X^{uL}_{tt}$.) 
Again, as we are only interested in the most 
stringent bound on the common factor   
$ke^{-\pi kR}$, the limit on $M^u_1$ from
the precise measurement of  $X^{uL}_{tt}$ is 
enough for closing the open window in the RS model with the 
SM fields off the wall.  

If we allow different quarks to be located at different points
of the fifth dimension (which could explain the fermion
mass hierarchy~\cite{gherghetta,flavour:hierarchy}) 
and then for non-diagonal $X$ corrections, 
one must wonder about possibly large FCNC and CP violation.
These can originate from the exchange of KK gauge bosons 
~\cite{gherghetta,Delgado:2000sv} and from mixing with KK fermions.
In the first case the effective scale must be very large (or as above, 
the first two families must be almost at the same location or 
in the region of equal coupling, $c_{L(R)} \gtsim 0. 5\ (\ltsim -0.5)$). 
We are interested, however, in the effects of quark mixing induced by 
the tower of vector-like quarks with different family location. 
In this case FCNC and CP violation are not too large 
due to the scaling of the $X_{ij}$ corrections with 
the quark masses $m_im_j$ (see Eqs. (\ref{xul1}-\ref{xdr1})). 
This is enough to suppress those effects below experimental 
limits if we require $X^{uL}_{tt}$ to agree with its SM value 
at LHC. Indeed, 
$|\Delta X _{ij} | \sim \frac{m_im_j}{m_t^2} |\Delta X^{uL}_{tt}| $,   
with $\Delta X^{uL}_{tt} = -m_t^2|(U_R^u)_{kt}|^2[\ \ ]_k$ and 
$\sum _{k=1}^{3}|(U_R^u)_{kt}|^2 = 1$ and $[\ \ ]_k$ positive. 
Then the limits above apply for some $k$ because not all mixing 
elements vanish. 
(Similar arguments apply for $W_{ij}$ but with small mixing angles 
(see Eqs. (\ref{wl1}, \ref{wr1})).)
This kind of behaviour has been usually assumed in SM extensions 
with vector-like quarks (see Ref.~\cite{Frampton:2000xi} and references 
there in). The models with extra dimensions realize naturally 
this scaling, introducing an infinite tower of exotic fermions with 
contributions dominated by the lightest states. 
A detailed study of FCNC and CP violation and their different origins 
will be presented elsewhere.

In conclusion, we have shown that if fermions are allowed to live in
the bulk of the RS model, mixing effects between the SM
fermions and the tower of their KK (vector-like) 
excitations can be important. In the conformal limit region, this
mixing gives 
 a strong constraint, even in
the absence of FCNC. At this point, other experimental tests
(including both electroweak precision measurements and direct collider
searches) provide smaller lower bounds from the decoupling of the
towers of gauge bosons and gravitons. Thus, the inclusion of quark mixing 
in the phenomenological study of models with extra dimensions
results in a significant improvement of the experimental constraints, 
allowing to  completely cover the parameter space up to scales
of the order of several TeV.    

\section*{Acknowledgements}
It is a pleasure to thank Ll. Ametller, A. Delgado, M. Masip, 
M. P\'{e}rez-Victoria, A. Pomarol, M. Quir\'{o}s and F. Zwirner
for discussions. JS thanks the
Dipartimento di Fisica Galileo Galilei and INFN Sezione di Padova for
their warm hospitality. This work has been
supported by CICYT, by Junta de
Andaluc\'{\i}a and by the European Union under contract
HPRN-CT-2000-00149. JS thanks MECD for financial support.

\end{document}